\documentclass[twocolumn,10pt]{IEEEtran}
\usepackage{algorithm,algorithmic,amsbsy,amsmath,amssymb,epsfig,bbm,mathrsfs,fancyhdr,fancyvrb,subfigure,url,color}

\usepackage{amsmath}
\usepackage{mathtools}
\usepackage{enumerate}
\usepackage{paralist}
\usepackage{graphicx}
\usepackage{epstopdf}
\graphicspath{{Figures/}}
\usepackage{subfigure}
\usepackage{cite}
\usepackage{array}
\usepackage{booktabs}
\usepackage{comment}
\usepackage{cases}
\usepackage{algorithm}
\usepackage{algorithmic}

\usepackage{amssymb}

\begin{document}

\title{Resource Allocation for Network-Integrated Device-to-Device Communications Using Smart Relays}

\author{Monowar Hasan and Ekram Hossain \\ Department of Electrical and Computer Engineering, University of Manitoba, Winnipeg, Canada
}

\date{}

\maketitle

\begin{abstract}
With increasing number of autonomous heterogeneous devices in future mobile networks, an efficient resource allocation scheme is required to maximize network throughput and achieve higher spectral efficiency.
In this paper, performance of network-integrated device-to-device (D2D) communication is investigated where D2D traffic is carried through relay nodes. An optimization problem is formulated for allocating radio resources to maximize end-to-end rate as well as conversing QoS requirements for cellular and D2D user equipment under total power constraint. Numerical results show that there is a distance threshold beyond which relay-assisted D2D communication significantly improves network performance when compared to direct communication between D2D peers.

\end{abstract}


\section{Introduction} \label{sec:intro}

Device-to-device (D2D) communication underlaying cellular network has recently been intensively discussed in standardization committee and academia. Reusing the LTE-A cellular resources, D2D communication enables wireless peer-to-peer services directly between user equipments (UEs)  which enhances spectrum utilization and improves cellular coverage. Possible usage cases for D2D communication are local voice and data services including \textit{content sharing} (i.e., exchanging photos, videos or documents through smart phones) and \textit{multiplayer gaming} \cite{d2d_example}. 

In the context of D2D communication, it becomes a crucial issue to set up reliable direct links between the UEs while satisfying quality-of-service (QoS) of traditional cellular UEs (CUEs) and D2D UEs in the network. Besides, interference to and from CUEs and poor propagation channel may limit the advantages of D2D communication in practical scenarios. In such cases, network assisted transmission through relays could efficiently enhance the performance of D2D communication when D2D-pairs are too far away from each other or the quality of D2D channel is not good enough for direct communication. 

In this paper, we consider relay-assisted D2D communication in LTE-A cellular networks where D2D-pairs are served by the relay node. We concentrate on the scenario in which potential D2D UEs are located near to each other (i.e, office blocks or university areas, concert sites etc.); however, the proximity and link condition may not be favorable for direct communication. Thanks to LTE-A Layer-3 (L3) relay featuring with self-backhauling configuration which makes it capable to perform operations similar to those of a base station (i.e., Evolved Node B [eNB] in  an LTE-A network). We formulate the resource allocation problem with an objective to maximizing the end-to-end rate (i.e., minimum achievable rate over two hops) for the UEs while maintaining the QoS (i.e., rate) requirements for cellular and D2D UEs under total power constraint at the relay node. The resource allocation problem turns out to be a mixed-integer non-linear programming (MINLP) problem and to make it tractable we relax it using the time-sharing strategy. The contribution of this paper is the analysis of network performance under relay-assisted D2D communication. The numerical results show that after a distance threshold relaying D2D traffic provides significant gain in achievable data rate.

The remainder of this paper is organized as follows. A review of related work is presented in Section \ref{sec:related_works}. Section \ref{sec:background} introduces LTE-A access methods and the relaying mechanisms. In Section \ref{sec:sys_model}, we present the system model and  formulate  the resource allocation problem. The permanence evaluation results are presented in Section \ref{sec:performance_eval} and finally we conclude the paper in Section \ref{sec:conclusion}  outlining possible future works.

\section{Related Work and Motivations} \label{sec:related_works}

Resource allocation in context of D2D communication for future generation OFDMA based wireless networks is one of the active areas of research. In \cite{zul-d2d}, a greedy heuristic based resource allocation scheme is devolved  for both uplink and downlink scenarios where a D2D-pair shares same resources with traditional user if the achieved SINR is greater than a threshold SINR.  A resource allocation scheme based on a column generation method is proposed in \cite{phond-d2d} to maximize the spectrum utilization by finding the minimum transmission length (i.e., time slots) for D2D links while protecting cellular users from interference and guaranteeing QoS. A distributed suboptimal joint mode selection and resource allocation scheme is proposed in \cite{d2d-ra-2} to reduce intracell and intercell interference. In \cite{xen-1}, authors consider relay selection and resource allocation for uplink scenarios with two classes of users having different (i.e., specific and flexible) rate requirements. The objective is to maximize system throughput by satisfying rate requirements for the rate-constraint users while confining the transmit power within power-budget. In \cite{d2d-rel-1}, performance (i.e., maximum ergodic capacity and outage probability) of cooperative relaying in relay-assisted D2D communication is investigated considering power constraints at eNB and numerical results show that multi-hop relaying lowered the outage probability. However, in \cite{zul-d2d,phond-d2d, d2d-ra-2, xen-1}, the effect of using relays in D2D communication is not studied. 

As a matter of fact, relaying mechanism explicitly in context of D2D communication has not been considered so far in the literature and most of the resource allocation schemes consider only one  D2D link. Taking the advantage of L3 relays supported by the 3GPP standard, we study the network performance of network-integrated D2D communication and show that relay-assisted D2D communication provides significant performance gain for long distance D2D links. A brief review of radio access and relaying mechanism in the LTE-A standard is provided next.

\section{Radio Access and Relaying in 3GPP LTE-A} \label{sec:background}

\subsection{Radio Access Methods in LTE-A Networks}

In the LTE-A radio interface, two consecutive time slots create a subframe where each timeslot spans 0.5 msec. Resources are allocated to UEs\footnote{By the term ``UE'', we refer to both cellular and D2D user equipments.} in units of resource blocks (RBs) over a subframe. Each RB occupies 1 slot (0.5 msec) in time domain and 180 KHz in frequency domain with subcarrier spacing of 15 KHz. The multiple access scheme for downlink (i.e., eNB/relay-to-UE) is OFDMA while the access scheme for uplink (i.e., UE-to-relay/eNB, relay-to-UE) is single carrier-FDMA (SC-FDMA). In general, SC-FDMA requires contiguous set of subcarrier allocation to UEs. Resource allocation in downlink supports both block-wise transmission \textit{(localized allocation)} and transmission on non-consecutive subcarriers \textit{(distributed allocation)}. For uplink transmission, current specification supports only localized resource allocation \cite{lte_book_1}.

\subsection{Relays in LTE-A Networks}

Relay node in LTE-A is wirelessly connected to radio access network through a donor eNB and serves UEs. Depending on the function, different relaying mechanisms used in LTE-A \cite{relay-book-1}. \textit{Layer 1 (L1)} relays act as repeaters, amplifying the input signal without and decoding/re-encoding. The L1 relays can either use the same carrier frequency (i.e., in-band relaying) or an orthogonal carrier frequency (i.e., out-of-band relaying). The main advantages of L1 relays are simplicity, cost-effectiveness, and low delay. However, with L1 relaying, noise and interference are also amplified and retransmitted. Hence, the SINR of the signal may deteriorate. 

\textit{Layer 2 (L2)} relays are also known as decode and forward (DF) relay which involves decoding the source signal at the relay node. The advantage of DF relays is that noise and interference do not propagate to the destination. However, a substantial delay occurs during the relaying operation. A L2 relay does not issue any scheduling information or any control signal (i.e., HARQ and channel feedback). Hence, an L2 relay cannot generate a complete cell and from a UE's perspective, it is only a part of donor cell. 

\textit{Layer 3 (L3)} relays with self-backhauling configuration performs the same operation as eNB except for lower transmit power and smaller cell size. It controls cell(s) and each cell has its own cell identity. The relay shall transmit its own control signals and UE shall receive scheduling information and HARQ feedback directly from the relay node.

When the link condition between D2D peers is poor or the distance is too far for direct communication, with the support of L3 relays, scheduling and resource allocation for D2D UE can be done in relay node and D2D traffic can be transmitted through relay. We refer to this as \textit{network-integrated D2D communication} which can be an alternative approach to provide better quality of service between distant D2D-links. In the next section, we describe the network configuration and present the formulation for resource allocation. 

\section{System Model and Problem formulation} \label{sec:sys_model}

\begin{figure}[!t]
\centering
\includegraphics[scale=0.61]{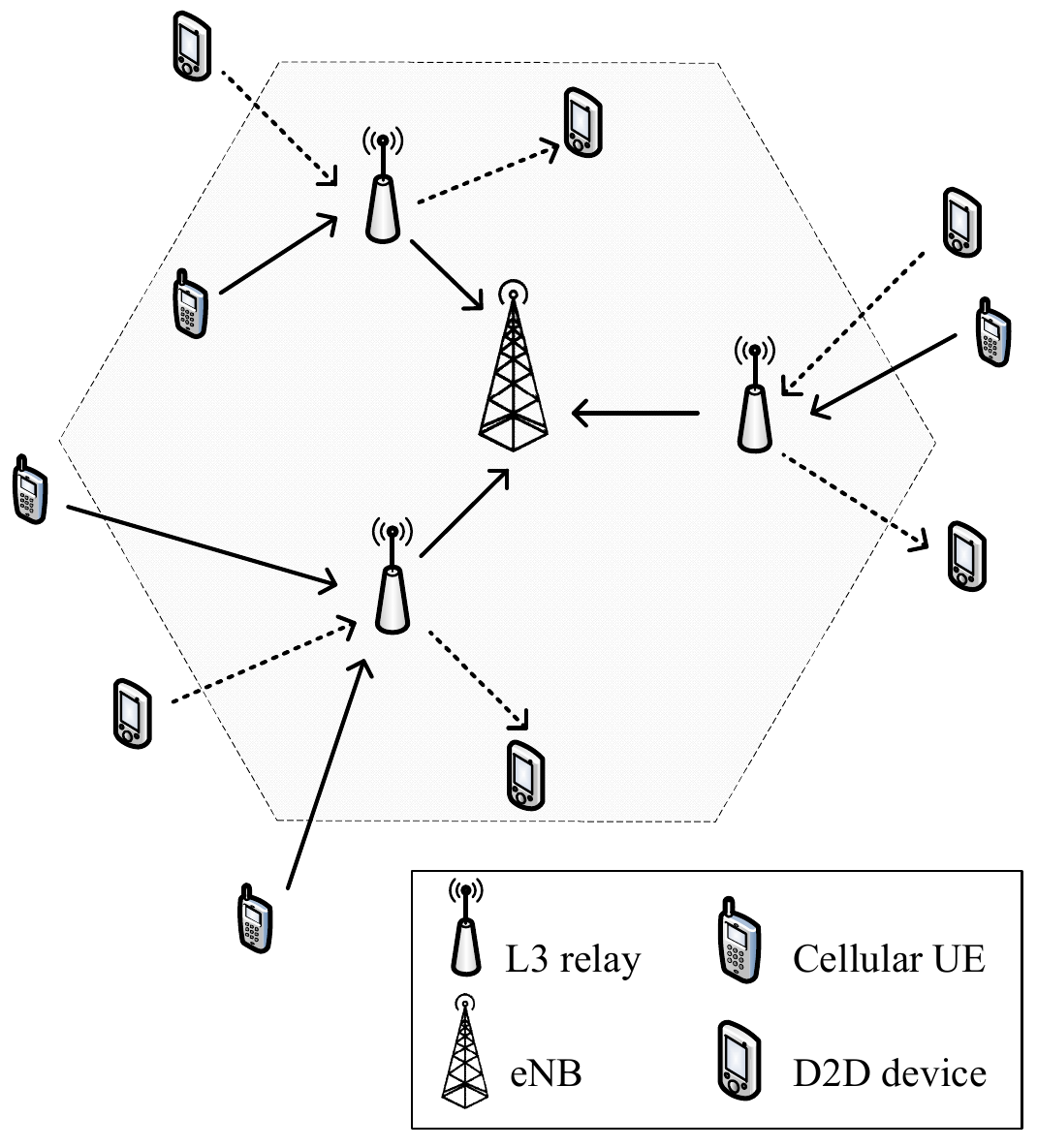}
\caption{A single cell with multiple relay nodes.} 
\label{fig:nw_diagram}
\end{figure}

 Let $\mathcal{L} = \lbrace 1, 2, \ldots, L \rbrace$ fixed-location L3 relays are available in the network as shown in Fig. \ref{fig:nw_diagram}. The CUEs and D2D-pairs  correspond to set $\mathcal{C}$ and $\mathcal{D}$, respectively, where the D2D-pairs are discovered during the D2D session setup. We consider localized resource allocation where system bandwidth is divided into $N$ RBs denoted by $\mathcal{N} = \lbrace 1, 2, \ldots, N \rbrace$. We assume that the CUEs are outside the coverage region of eNB and/or having bad channel condition, and therefore, CUE-eNB communications need to be supported by the relays. Besides, the direct communication between two D2D UEs could be unfavourable due to long distance and/or poor link condition, and therefore, requires the assistance of a relay node. The UEs (i.e., both cellular and D2D) assisted by relay $l$ are denoted by $u_l$. The set of UEs assisted by relay $l$ is $\mathcal{U}_l$ such that $\mathcal{U}_l \subseteq \lbrace \mathcal{C} \cup \mathcal{D} \rbrace, \forall l \in \mathcal{L}$; $\bigcup_l \mathcal{U}_l = \lbrace \mathcal{C} \cup \mathcal{D} \rbrace$ and  $\bigcap_l \mathcal{U}_l = \varnothing$.  According to our system model, taking the advantage of an L3 relay, scheduling and resource allocation is performed in the relay node to reduce overload at the eNB.

We define $h_{i,j}^{n}$ the link gain between the link $i$ and $j$ over RB $n$. The unit power ${\rm SINR}$ for the link between UE $u_l \in \mathcal{U}_l$ and relay $l$ using RB $n$ in the first hop is as follows:
\begin{equation}
\label{eq:SINR_1}
\gamma_{u_l, l, 1}^n = \frac{h_{u_l, l}^n}{\displaystyle \sum_{\forall u_j \in \mathcal U_j, j \neq l, j \in \mathcal{L}} P_{u_j, j}^n h_{u_j, l}^n + N_0 B_{RB}}.
\end{equation}
The unit power ${\rm SINR}$ for the link between relay $l$ and eNB  for CUE (i.e.,  $u_l \in \lbrace \mathcal{C} \cap \mathcal{U}_l \rbrace$) in the second hop is given by
\begin{equation}
\label{eq:SINR_2}
\gamma_{l, u_l, 2}^n = \frac{h_{l, eNB}^n}{\displaystyle \sum_{\forall u_j \in  \mathcal{U}_j, j \neq l, j \in \mathcal{L}} P_{j, u_j}^n h_{j, eNB}^n + N_0 B_{RB}}.
\end{equation}

Similarly, the unit power ${\rm SINR}$ for the link between relay $l$ and receiving D2D UE for the D2D-pair (i.e., $u_l \in \lbrace \mathcal{D} \cap \mathcal{U}_l \rbrace$) in the second hop can be written as
\begin{equation}
\label{eq:SINR_3}
\gamma_{l, u_l, 2}^n = \frac{h_{l, u_l}^n}{\displaystyle \sum_{\forall u_j \in  \mathcal{U}_j , j \neq l, j \in \mathcal{L}} P_{j, u_j}^n h_{j, u_l}^n + N_0 B_{RB}   }
\end{equation}
where $P_{i, j}^n$ is the power assigned between link $i$ and $j$ over RB $n$, $B_{RB}$ is bandwidth of RB, and $N_0$ denotes thermal noise.  $h_{l, eNB}^n$ is the gain between relay-eNB link; $h_{l, u_l}^n$ is the gain between relay $l$ and receiving D2D UE for the D2D-pair $u_l$.

The achievable data rate for the UE $u_l$  in the first hop can be expressed as, $r_{u_l, 1}^n = B_{RB} \log_2 ( 1 + P_{u_l, l}^n \gamma_{u_l, l, 1}^n)$. Similarly, the achievable data rate in the second hop is as follows: $r_{u_l, 2}^n = 
B_{RB} \log_2 ( 1 + P_{l, u_l}^n \gamma_{l, u_l, 2}^n)$. Note that,
for the CUE (i.e., UEs $\in \lbrace \mathcal{C} \cap \mathcal{U}_l \rbrace$), the ${\rm SINR}$ in the second hop is calculated from (\ref{eq:SINR_2}); on the other hand, the ${\rm SINR}$ for the D2D UEs (i.e., UEs $\in \lbrace \mathcal{D} \cap \mathcal{U}_l \rbrace$) is calculated from (\ref{eq:SINR_3}).

The end-to-end data rate on RB $n$ for the UE $u_l$ is the minimum achievable data rate over two hops, i.e.,   
\begin{equation}
\label{eqn:e2e_rate}
R_{u_l}^n =  \frac{1}{2} \min \left\lbrace  r_{u_l, 1}^n , r_{u_l, 2}^n    \right\rbrace.
\end{equation}

\subsection{Resource Block and Power Allocation in Relay Nodes}

The objective of resource allocation (i.e., RB and transmit power allocation) is to specify for each relay, the RB and power level assignment to the UEs which maximizes the system capacity defined as the minimum achievable data rate over two hops. 

Let the maximum allowable transmit power for UE (relay) is $P_{u_l}^{max}$ ($P_l^{max}$). The RB allocation indicator is denoted by binary decision variable $x_{u_l}^n \in \lbrace 0, 1\rbrace$, where $x_{u_l}^n = 1$ if RB $n$ is assigned to UE $u_l$ and $0$, otherwise. The same RB(s) will be used by relay in the second hop and $\displaystyle R_{u_l} = \sum_{n =1}^N  x_{u_l}^n R_{u_l}^n$ denotes the achievable sum-rate over allocated RB(s). $Q_{u_l}$ denotes the QoS (rate) requirements for a UE $u_l$. The resource allocation problem for each relay $l \in \mathcal{L}$ can be formulated as follows:

\begin{subequations}
\setlength{\arraycolsep}{0.0em}
\begin{eqnarray}
\mathbf{(P1)} ~ \underset{x_{u_l}^n, P_{u_l, l}^n, P_{l, u_l}^n}{\operatorname{max}} ~ \sum_{u_l \in \mathcal{U}_l } &&  \sum_{n =1}^N   x_{u_l}^n R_{u_l}^n  \nonumber \\
\text{subject to} \quad \sum_{u_l \in \mathcal{U}_l} x_{u_l}^n ~ &&{\leq} ~ 1,  \quad ~~~\forall n \in \mathcal{N} \label{eq:con-bin} \\
\quad \sum_{n =1}^N x_{u_l}^n P_{u_l, l}^n ~ &&{\leq} ~ P_{u_l}^{max}, ~ \forall u_l \in \mathcal{U}_l \label{eq:con-pow-ue} \\
\quad \sum_{u_l \in \mathcal{U}_l } \sum_{n =1}^N x_{u_l}^n P_{l, u_l}^n ~ &&{\leq} ~ P_l^{max}  \label{eq:con-pow-rel} \\
\quad \sum_{u_l \in \mathcal{U}_l } x_{u_l}^n P_{u_l, l}^n h_{{u_l}_{ref}, l, 1}^n ~ &&{\leq} ~ I_{th, 1}^n,  ~~\forall n \in \mathcal{N} \label{eq:con-intf-1}\\
\quad \sum_{u_l \in \mathcal{U}_l } x_{u_l}^n  P_{l, u_l}^n h_{{u_l}_{ref}, l, 2}^n ~ &&{\leq} ~ I_{th, 2}^n, ~~\forall n \in \mathcal{N} \label{eq:con-intf-2} \\
\quad R_{u_l}  ~ &&{\geq} ~ Q_{u_l}, \quad \forall u_l \in \mathcal{U}_l  \label{eq:con-QoS-cue}\\
\quad P_{u_l, l}^n ~ \geq ~ 0, ~~ P_{l, u_l}^n ~ &&{\geq} ~ 0,  \quad ~~~\forall n \in \mathcal{N}, u_l \in \mathcal{U}_l \label{eq:con-pow-0}
\end{eqnarray}
\setlength{\arraycolsep}{5pt}
\end{subequations}
where $R_{u_l}^n = \frac{1}{2} \min \left\lbrace \begin{smallmatrix}
B_{RB} \log_2 ( 1 + P_{u_l, l}^n \gamma_{u_l, l, 1}^n),\\ 
B_{RB} \log_2 ( 1 + P_{l, u_l}^n \gamma_{l, u_l, 2}^n)   
\end{smallmatrix} \right\rbrace$, ${\rm SINR}$  for the first hop, \[\gamma_{u_l, l, 1}^n = \frac{h_{u_l, l}^n}{\displaystyle \sum_{\forall u_j \in \mathcal U_j, j \neq l, j \in \mathcal{L}} x_{u_j}^n P_{u_j, j}^n h_{u_j, l}^n + N_0 B_{RB}  }\] and ${\rm SINR}$ for the second hop, 
\begin{numcases}{\gamma_{l, u_l, 2}^n = }
\frac{h_{l, eNB}^n}{\displaystyle \sum_{\forall u_j \in  \mathcal{U}_j, j \neq l, j \in \mathcal{L}} x_{u_j}^n P_{j, u_j}^n h_{j, eNB}^n + N_0 B_{RB}}, & \nonumber \\
& \hspace{-9em}  $u_l \in \lbrace \mathcal{C} \cap \mathcal{U}_l \rbrace$ \nonumber \\
\frac{h_{l, u_l}^n}{\displaystyle \sum_{\forall u_j \in  \mathcal{U}_j , j \neq l, j \in \mathcal{L}} x_{u_j}^n  P_{j, u_j}^n h_{j, u_l}^n + N_0 B_{RB}}, &  \nonumber \\
& \hspace{-9em} $u_l \in \lbrace \mathcal{D} \cap \mathcal{U}_l \rbrace$. \nonumber
\end{numcases}

With the constraint in (\ref{eq:con-bin}), each RB is assigned to only one UE. Under the constraints in (\ref{eq:con-pow-ue}) and (\ref{eq:con-pow-rel}), the transmit power is limited by maximum power budget. (\ref{eq:con-intf-1}) and (\ref{eq:con-intf-2}) constraint the amount of interference introduced to other relays and receiving D2D UEs in first and second hop, respectively, to be less than some threshold. Constraint (\ref{eq:con-QoS-cue}) ensures the minimum QoS requirements for the CUE and D2D UEs. The constraint in (\ref{eq:con-pow-0}) is the non-negativity condition of transmit power.

Similar to \cite{ref_user}, we adopt the concept of reference node. For example, in the first hop, each UE associated with relay node $l$ chooses from among the neighbouring relays having the highest channel gain according to (\ref{eq:ref_user1}) and allocates the power level considering the interference threshold. Similarly, in the second hop, transmit power for each relay $l$  will be adjusted accordingly considering interference introduced to receiving D2D UEs (associated with neighbouring relays) according to (\ref{eq:ref_user2}).
\begin{subequations}
\setlength{\arraycolsep}{0.0em}
\begin{eqnarray}
\hspace{-1.5em} h_{{u_l}_{ref}, l, 1}^n &&= \underset{j}{\operatorname{argmax}} ~ h_{u_l , j}^n;  u_l \in \mathcal{U}_l, j \neq l, j \in \mathcal{L}. \label{eq:ref_user1}\\
\hspace{-1.5em} h_{{u_l}_{ref}, l, 2}^n &&= \underset{u_j}{\operatorname{argmax}} ~ h_{l , u_j}^n;   j \neq l, j \in \mathcal{L},  u_j \in \lbrace \mathcal{D} \cap \mathcal{U}_j \rbrace. \label{eq:ref_user2}
\end{eqnarray}
\setlength{\arraycolsep}{5pt}
\end{subequations}

From (\ref{eqn:e2e_rate}), the maximum rate for UE $u_l$ over RB $n$ is achieved when $P_{u_l, l}^n \gamma_{u_l, l, 1}^n = P_{l, u_l}^n \gamma_{l, u_l, 2}^n$. Therefore, power allocated to relay node for UE $u_l$ can be expressed as a function of power at UE as $P_{l, u_l}^n  = \frac{\gamma_{u_l, l, 1}^n}{\gamma_{l, u_l, 2}^n}P_{u_l, l}^n$ and the rate of UE $u_l$ over RB $n$, 

\begin{equation}
R_{u_l}^n = \frac{1}{2} B_{RB}  \log_2 \left( 1 + P_{u_l, l}^n \gamma_{u_l, l, 1}^n \right).
\end{equation}  

The optimization problem $\mathbf{P1}$ is a mixed-integer non-linear program (MINLP) which is computationally intractable and very complex to solve. A common approach in literature is to relax the constraint that an RB is used by only one UE using time-sharing factor \cite{relax-con-1}. Thus $x_{u_l}^n \in (0,1]$ is represented as the sharing factor where each $x_{u_l}^n$ denotes the portion of time that RB $n$ is assigned to UE $u_l$ and satisfies $\displaystyle \sum_{u_l \in \mathcal{U}_l} x_{u_l}^n \leq 1, ~\forall n$. Besides, we introduce a new variable $S_{u_l, l}^n = x_{u_l}^n P_{u_l,l}^n$ which denotes the actual transmit power of UE $u_l$ on RB $n$ \cite{time-share-1}. Then the relaxed problem can be reformulated as follows:  
\begin{subequations}
\setlength{\arraycolsep}{0.0em}
\begin{eqnarray}
 \mathbf{(P2)} ~ 
\underset{x_{u_l}^n, S_{u_l, l}^n}{\operatorname{max}} ~ \sum_{u_l \in \mathcal{U}_l } \sum_{n =1}^N   \frac{1}{2}  x_{u_l}^n   B_{RB}  \log_2  && \left(  1 +   \frac{S_{u_l, l}^n \gamma_{u_l, l, 1}^n}{x_{u_l}^n} \right)    \nonumber \\
\text{subject to} \quad \sum_{u_l \in \mathcal{U}_l} x_{u_l}^n ~ &&{\leq} ~ 1, \quad \forall n \hspace{-5em} \label{eq:con-bin-relx} \\
\quad \sum_{n =1}^N S_{u_l, l}^n ~ &&{\leq} ~ P_{u_l}^{max}, ~\forall u_l  \label{eq:con-pow-ue-relx} \\
\quad \sum_{u_l \in \mathcal{U}_l } \sum_{n =1}^N \frac{\gamma_{u_l, l, 1}^n}{\gamma_{l, u_l, 2}^n} S_{u_l, l}^n ~ &&{\leq} ~ P_l^{max} \label{eq:con-pow-rel-relx} \\
\quad \sum_{u_l \in \mathcal{U}_l } S_{u_l, l}^n h_{{u_l}_{ref}, l, 1}^n ~ &&{\leq} ~ I_{th, 1}^n, ~~ \forall n \label{eq:con-intf-1-relx}\\
\quad \sum_{u_l \in \mathcal{U}_l } \frac{\gamma_{u_l, l, 1}^n}{\gamma_{l, u_l, 2}^n} S_{u_l, l}^n h_{{u_l}_{ref}, l, 2}^n ~ &&{\leq} ~ I_{th, 2}^n, ~~\forall n \label{eq:con-intf-2-relx} \\
\quad \sum_{n=1}^N \frac{1}{2}  x_{u_l}^n B_{RB} \log_2 \left(  1 + \frac{S_{u_l, l}^n \gamma_{u_l, l, 1}^n}{x_{u_l}^n} \right)  ~ &&{\geq} ~ Q_{u_l},  ~~ \forall u_l   \label{eq:con-QoS-cue-relx}\\
\quad S_{u_l, l}^n  ~ &&{\geq} ~ 0,  ~~\forall n, u_l. \label{eq:con-pow-0-relx}
\end{eqnarray}
\setlength{\arraycolsep}{5pt}
\end{subequations}

The duality gap of any optimization problem satisfying time sharing condition is negligible as the number of subcarrier  becomes significantly large. Since our optimization problem satisfies the time-sharing condition, the solution of the relaxed problem is asymptotically optimal \cite{large-rb-dual}. The optimization problem $\mathbf{P2}$ is convex; the objective function is concave, constraint (\ref{eq:con-QoS-cue-relx}) is convex and all the remaining constraints are affine. Therefore this problem can be solved by the interior point method \cite{book-boyd}. 

To observe the nature of power allocation for a UE, we use Karush-Kuhn-Tucker (KKT) optimality and define the following Lagrangian function:

\begin{flalign}
    &\mathbb{L} (\mathbf{x}, \mathbf{S},\boldsymbol{\mu},\boldsymbol{\rho}, \nu_l, \boldsymbol{\psi}, \boldsymbol{\varphi}, \boldsymbol{\lambda}) =  \nonumber \\ 
&- \sum_{u_l \in \mathcal{U}_l} \sum_{n = 1}^{N} \tfrac{1}{2} x_{u_l}^n B_{RB} \log_2 \left(1+ \tfrac{S_{u_l, l}^n \gamma_{u_l, l, 1}^n}{x_{u_l}^n}\right) \nonumber \\
&+ \sum_{n = 1}^{N}  \mu_n  \left(\sum_{u_l \in \mathcal{U}_l} x_{u_l}^n -1 \right)  + \sum_{u_l \in \mathcal{U}_l}  \rho_{u_l} \left( \sum_{n = 1}^{N} S_{u_l, l}^n - P_{u_l}^{max} \right) \nonumber \\
&+ \nu_l \left( \sum_{u_l \in \mathcal{U}_l } \sum_{n =1}^N \tfrac{\gamma_{u_l, l, 1}^n}{\gamma_{l, u_l, 2}^n} S_{u_l, l}^n - P_l^{max} \right) \nonumber \\
&+ \sum_{n = 1}^{N}  \psi_n  \left( \sum_{u_l \in \mathcal{U}_l } S_{u_l, l}^n h_{{u_l}_{ref}, l, 1}^n ~ - I_{th, 1}^n \right) \nonumber \\
&+ \sum_{n = 1}^{N}  \varphi_n  \left(\sum_{u_l \in \mathcal{U}_l } \tfrac{\gamma_{u_l, l, 1}^n}{\gamma_{l, u_l, 2}^n} S_{u_l, l}^n h_{{u_l}_{ref}, l, 2}^n - I_{th, 2}^n \right) \nonumber \\
&+ \sum_{u_l \in \mathcal{U}_l }  \lambda_{u_l} \left(Q_{u_l} - \sum_{n=1}^N \tfrac{1}{2}  x_{u_l}^n B_{RB} \log_2 \left(  1 + \tfrac{S_{u_l, l}^n \gamma_{u_l, l, 1}^n}{x_{u_l}^n} \right) \right)  \nonumber \\
\label{eq: lagrange-1} 
\end{flalign}
where $\boldsymbol{\lambda}$ is the vector of Lagrange multipliers associated with individual QoS requirements for cellular and D2D UEs. Similarly, $\boldsymbol{\mu},\boldsymbol{\rho}, \nu_l, \boldsymbol{\psi}, \boldsymbol{\varphi}$ are the Lagrange multipliers for constraint (\ref{eq:con-bin-relx})-(\ref{eq:con-intf-2-relx}). Differentiating (\ref{eq: lagrange-1}) with respect to $S_{u_l, l}^n$, we obtain the following power allocation for UE $u_l$ over RB $n$:
\begin{equation}
\label{eq:power-alloc}
P_{u_l, l}^n = \frac{S_{u_l,l}^n}{x_{u_l}^n} = \left[\delta - \frac{1}{ \gamma_{u_l,l, 1}^n}\right]^+
\end{equation} 
where $\delta = \frac{\tfrac{1}{2} B_{RB} \frac{(1 + \lambda_{u_l})}{\ln 2}}{\rho_{u_l} + \frac{\gamma_{u_l,l, 1}^n}{\gamma_{l, u_l,2}^n} \nu_l + h_{{u_l}_{ref}, l, 1}^n \psi_{n} + \frac{\gamma_{u_l, l, 1}^n}{\gamma_{l, u_l, 2}^n} h_{{u_l}_{ref}, l, 2}^n \varphi_{n} } $ and $[\varepsilon]^+ = \max(\varepsilon,0)$, which is a multi-level water filling allocation \cite{time-share-1}.

\subsection{Semi-distributed Resource Allocation Algorithm}


Each relay in the network independently allocates resources to its associated UEs. Based on the mathematical formulation in the previous section, the overall resource allocation algorithm is shown in \textbf{Algorithm \ref{alg:rec_alloc}}.

\begin{algorithm}
\caption{Joint RB and power allocation algorithm}
\label{alg:rec_alloc}
\begin{algorithmic}[1]   

\STATE UEs measure interference level from previous time slot and inform the respective relays.

\STATE  Each relay $l \in \mathcal{L}$ obtains the channel state information
among all relays $j; j \neq l, j \in \mathcal{L}$ and to its scheduled UEs $\forall u_j \in \mathcal{U}_j; j \neq l, j \in \mathcal{L}$. 

\STATE  For each relay and its associated UEs, obtain the reference node for the first and second hops according to (\ref{eq:ref_user1}) and (\ref{eq:ref_user2}).

\STATE  Solve the optimization problem $\mathbf{P2}$ for each relay independently to obtain RB and power allocation.

\STATE  Allocate resources (i.e., RB and transmit power) to associated UEs for each relay and calculate average achievable data rate.

\end{algorithmic}
\end{algorithm}

Since the L3 relays can perform the same operation as an eNB, these relays can communicate using the X2 interface \cite{lte_arch} defined in the 3GPP LTE-A standard. Therefore, in the proposed algorithm, the relays can obtain the channel state information through inter-relay message passing without increasing signalling overhead at the eNB.     
  
\section{Performance evaluation} \label{sec:performance_eval}

\subsection{Parameters}

The performance results for the resource allocation schemes obtained
by a simulator written in MATLAB. In order to measure channel gain, we consider both distant-dependent path-loss and shadow fading; and the channel is assumed to be frequency-selective and experience Rayleigh fading. In particular, we consider realistic 3GPP propagation environment\footnote{Any other propagation model for D2D communication can be used for the proposed resource allocation method.} presented in \cite{relay-book-2}. For example, UE-relay (and relay-D2D) link follows the following path-loss equation: $PL_{u_l,l}(\ell)_{[dB]} = 103.8+ 20.9 \log(\ell) + L_{su} + 10 \log(\phi)$, where $\ell$ is the distance between UE and relay in kilometer; $L_{su}$ accounts for shadow fading and is modelled as a log-normal random variable, and $\phi$ is an exponentially distributed random variable which represents Rayleigh fading. Similarly, the path-loss equation for relay-eNB link is expressed as: $PL_{l,eNB}(\ell)_{[dB]} = 100.7 + 23.5 \log(\ell) + L_{sr} + 10 \log(\phi)$, where $L_{sr}$ is a log-normal random variable accounting for shadow fading. The simulation parameters and assumptions used for obtaining the numerical results are listed in Table \ref{tab:sim_param}. 

\begin{table}[!t]
\renewcommand{\arraystretch}{1.3}
\caption{Simulation Parameters}
\label{tab:sim_param}
\centering
\begin{tabular}{c|c}
\hline
\bfseries Parameter & \bfseries Values\\
\hline\hline
Carrier frequency & $2.35$ GHz \\
System bandwidth & $2.5$ MHz \\
Total number of RBs for each relay & $13$ \\
Relay cell radius & $200$ meter\\
Distance between relay and eNB & $125$ meter\\
Minimum distance between UE and relay & $10$ meter\\
Total power available at each relay & $30$ dBm \\
Total power available at UE & $23$ dBm \\
Noise power spectral density & $-174$ dBm/Hz \\
Shadow fading standard deviation in relay-eNB link & $6$ dB\\
Shadow fading standard deviation in UE-relay link & $10$ dB\\
Rate requirement for cellular UEs & $100$ bps \\
Rate requirement for D2D UEs & $200$ bps \\
\hline
\end{tabular}
\end{table}

\begin{figure}[!t]
\centering
\includegraphics[scale=0.57]{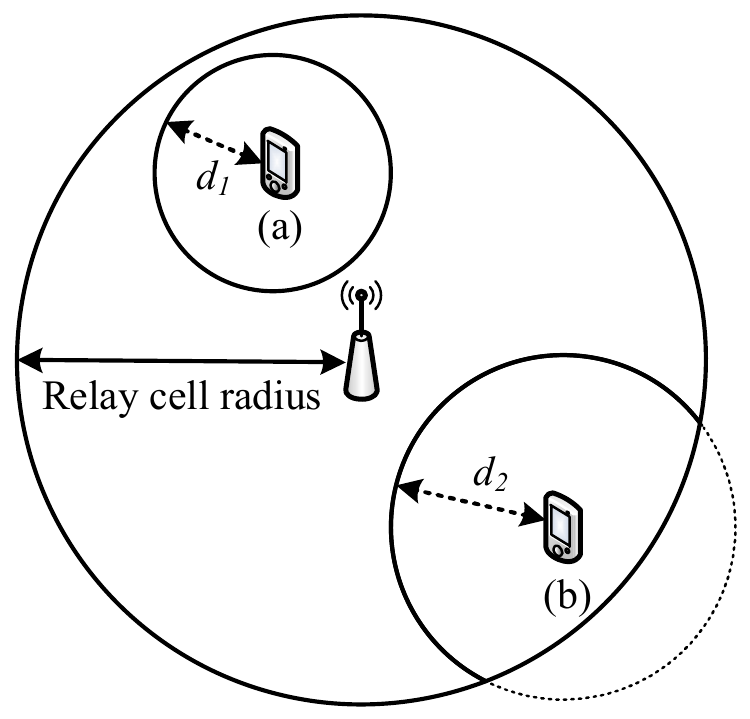}
\caption{Distribution of any D2D-pairs for two cases: (a) other UE of the D2D-pair is distributed anywhere on the edge of the circle with radius $d_1$; (b) other UE of the D2D-pair having distance $d_2$ between them is uniformly distributed on the solid arc.} 
\label{fig:d2d_position}
\end{figure}

We simulate a single three-sectored cellular network in an rectangular
area of 700m $\times$ 700m, where the eNB is located in the centre of the cell and three relays are deployed in the network, i.e., one relay in each sector. The CUEs and the transmitter UE of a D2D-pair are uniformly distributed within the radius of the relay cell. The other UE of the D2D-pair is distributed uniformly in the overlapping area of the relay radius and a circle centred at the first D2D UE as shown in Fig. \ref{fig:d2d_position}. The circle radius which gives the maximum distance between UEs in a D2D-pair is varied as a simulation parameter. The numerical results are averaged over different realizations of simulation scenarios (i.e., UE locations and channel gains).

\subsection{Numerical Results}

In order to study network performance in presence of the L3 relay, we compare the performance of the proposed scheme with a \textit{reference scheme} \cite{zul-d2d} in which an RB allocated to CUE can be shared with at most one D2D-link. D2D UE shares the same RB(s) (allocated to CUE by solving optimization problem) and communicate directly between peers without relay only if the QoS requirements for both CUE and D2D UE are satisfied.

\subsubsection{Achievable data rate vs. distance between D2D-links}

\begin{figure}[!t]
\centering
\includegraphics[scale=0.55]{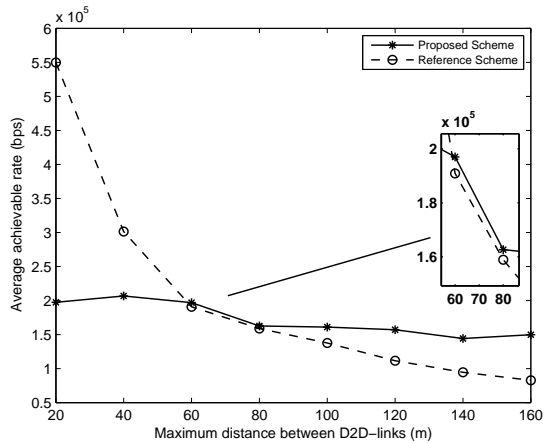}
\caption{Average achievable data rate with varying distance; number of CUE, $|\mathcal{C}| = 15$ (i.e., 5 CUEs assisted by each relay), number of D2D-pair,  $|\mathcal{D}| = 9$ (i.e., 3 D2D-pair assisted by each relay) and interference threshold -70 dBm.} 
\label{fig:rate_05_03}
\end{figure}

In Fig. \ref{fig:rate_05_03},  we illustrate the average achievable data rate $\bar{R}$ for D2D UEs which is calculated as $\bar{R} =  \frac{ \displaystyle \sum_{u \in \mathcal{D}} R_{u}^{ach}}{|\mathcal{D}|}$, where $R_{u}^{ach}$ is the achievable data rate for UE $u$ and $|\cdot|$ denotes set cardinality. Although the reference scheme outperforms when the distance between D2D-link is closer (i.e., $d < 60\text{m}$); our proposed algorithm can greatly increase the data rate especially when the distance increases. This is due to the fact that when the distance is higher, the performance of direct communication deteriorates due to poor propagation medium. Besides, when the D2D UEs share resources with only one CUE, the spectrum may not utilize efficiently and decreases the achievable rate. Consequently, the gap between the achievable rate of our proposed algorithm and that of the reference scheme becomes wider when the distance increases.

\begin{figure*}[h t b]
\centering
\subfigure[]{\includegraphics[scale=0.55]{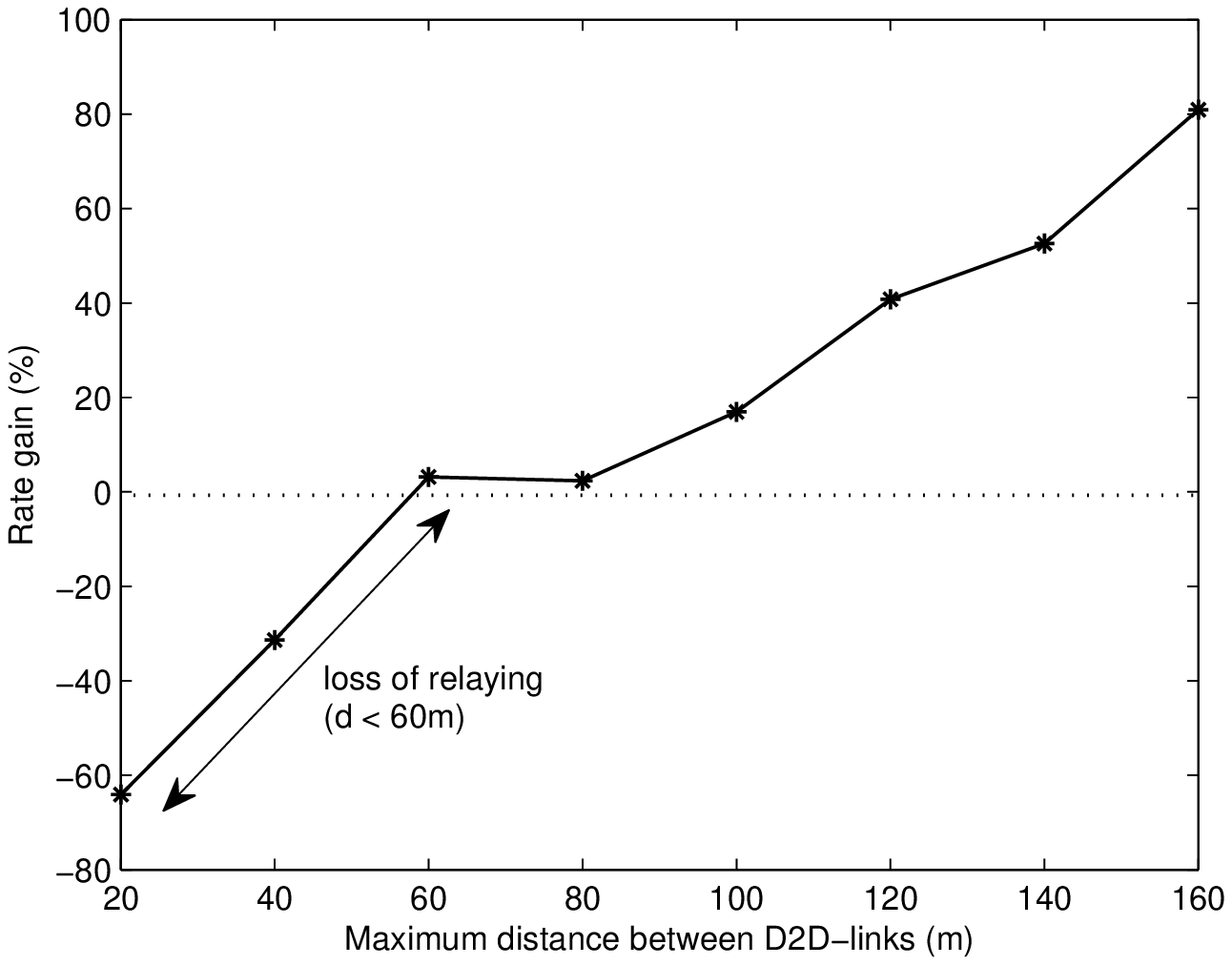}%
\label{fig:gain_05_03}} 
\hfil 
\subfigure[]{\includegraphics[scale=0.55]{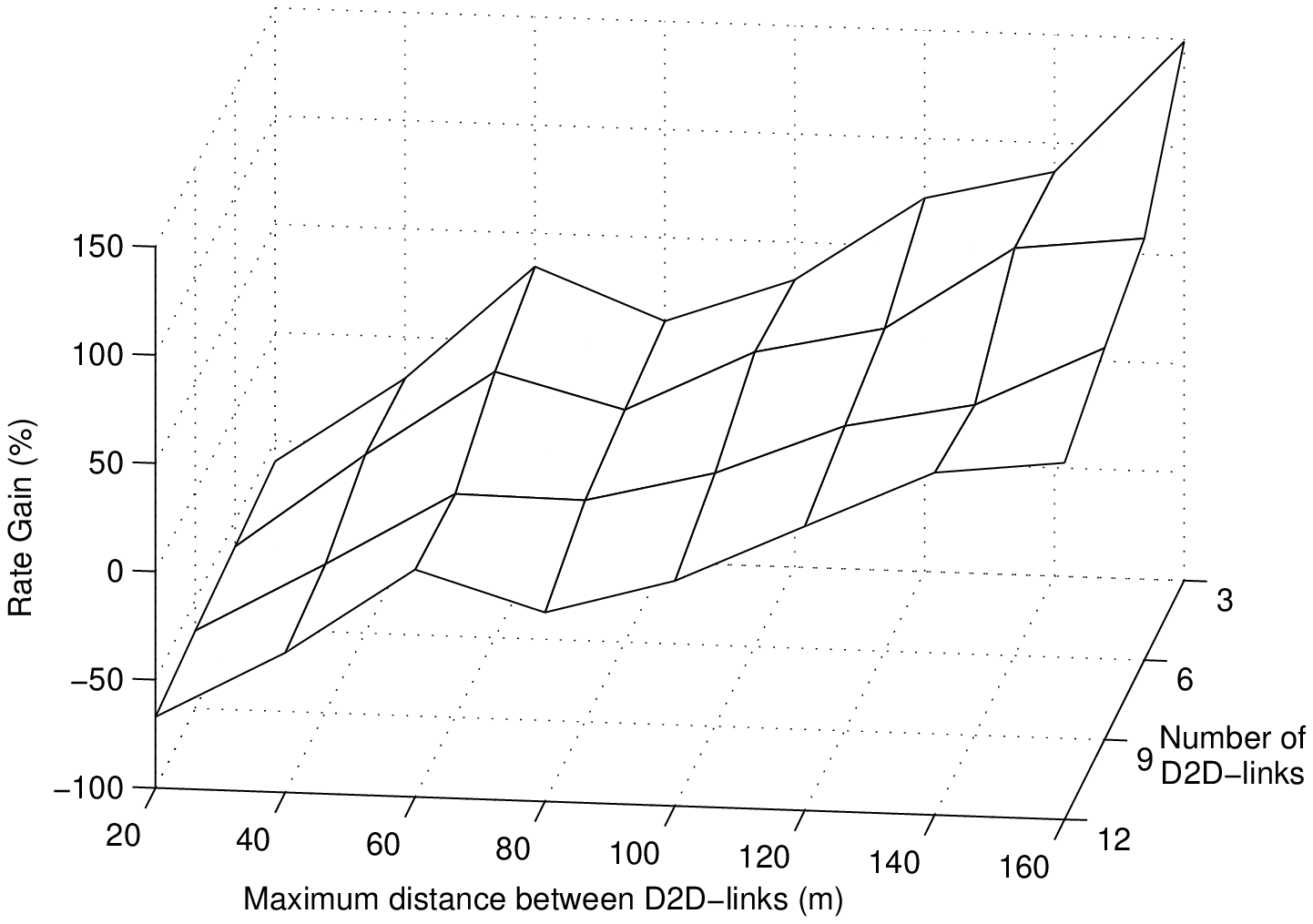}%
\label{fig:gain_dif}}
\caption{Gain in aggregated achievable data rate with varying distance (for $|\mathcal{C}| = 15$, interference threshold -70 dBm): (a) 3 D2D-pairs assisted by each relay (i.e., $|\mathcal{D}| = 9$);  (b) number of D2D-pairs varies from 1 to 4 UE(s)/relay (i.e., $|\mathcal{D}| = 3,6,9,12$). There is a critical distance $d$ (i.e., $d \approx 60 \text{m}$ here), beyond which relaying provides significant performance gain.}
\label{fig:rate_gain}
\end{figure*}

\subsubsection{Rate gain vs. distance between D2D-links}

Fig. \ref{fig:gain_05_03} depicts the rate gain in terms of  aggregated achievable rate for the UEs. We calculate the gain as follows: $Rate ~gain = \frac{R_{prop} - R_{ref}}{R_{ref}} \times 100 \% $ where $R_{prop}$ and  $R_{ref}$ is the aggregated rate for the UEs in proposed and reference scheme, respectively. It is observed from the figure that, with the increasing distance between D2D-links our proposed scheme provides significant gain in terms of achievable data rate. To observe the effect of gain in different network realization we vary the number of D2D UE in Fig. \ref{fig:gain_dif}. It is clear from figure that irrespective of the number of D2D UEs in the network, our proposed scheme provides considerable rate gain for distant D2D-pairs. 


\section{Conclusion} \label{sec:conclusion}

We have provided a mathematical formulation for resource allocation and analyzed network performance of relay-assisted D2D communication. The performance evaluation results have shown that relay-assisted D2D communication is beneficial to provide higher rate for distant D2D-links. Along with the rate requirements, it can be possible to measure additional QoS parameters (i.e., delay) for observing network performance properly by other mathematical tools (i.e., queuing models). Besides when the perfect channel knowledge and the information about number of active UEs are not available, the effects of uncertainties in the system parameter need to be considered by using a robust optimization formulation. These issues will be explored in our future works.


%


\bibliographystyle{IEEEtran}


\end{document}